\documentclass[twocolumn,aps,prl,superscriptaddress]{revtex4-2}
\usepackage{amssymb}
\usepackage{amsmath,bm}
\usepackage{graphicx}
\usepackage[usenames, dvipsnames]{xcolor}
\usepackage{pslatex}
\usepackage[colorlinks,linkcolor=blue,urlcolor=blue,citecolor=blue]{hyperref}

\allowdisplaybreaks[4]

\begin{document} 

\title{Chronoscoping Anisotropic Flow in the Quark-Gluon Plasma with Dileptons}

\author{Fu Ma} 
\affiliation{Key Laboratory of Nuclear Physics and Ion-beam Application (MOE), Institute of Modern Physics, Fudan University, Shanghai 200433, China}
\affiliation{Shanghai Research Center for Theoretical Nuclear Physics, NSFC and Fudan University, Shanghai 200438, China}

\author{Wen-Hao Zhou}
\email{Contact author: zhouwenhao@xaau.edu.cn}
\affiliation{Faculty of Science, Xihang University, Xi'an 710077, China}

\author{Kai-Jia Sun}
\email{Contact author: kjsun@fudan.edu.cn}
\affiliation{Key Laboratory of Nuclear Physics and Ion-beam Application (MOE), Institute of Modern Physics, Fudan University, Shanghai 200433, China}
\affiliation{Shanghai Research Center for Theoretical Nuclear Physics, NSFC and Fudan University, Shanghai 200438, China}

\author{Chi Yang}
\email{Contact author: chiyang@sdu.edu.cn}
\affiliation{Key Laboratory of Particle Physics and Particle Irradiation (MOE), Institute of Frontier and Interdisciplinary Science, Shandong University, Qingdao 266237, China}

\author{Che Ming Ko}
\email{Contact author: ko@comp.tamu.edu}
\affiliation{Cyclotron Institute and Department of Physics and Astronomy, Texas A\&M University, College Station, Texas 77843, USA}

\author{Yu-Gang Ma}
\email{Contact author: mayugang@fudan.edu.cn}
\affiliation{Key Laboratory of Nuclear Physics and Ion-beam Application (MOE), Institute of Modern Physics, Fudan University, Shanghai 200433, China}
\affiliation{Shanghai Research Center for Theoretical Nuclear Physics, NSFC and Fudan University, Shanghai 200438, China}
\affiliation{School of Physics, East China Normal University, Shanghai 200241, China}

\author{Zhangbu Xu}
\email{Contact author: zxu22@kent.edu}
\affiliation{Physics Department, Kent State University, Kent, OH 44242, USA}
\affiliation{Physics Department, Brookhaven National Laboratory, Upton, NY 11973, USA}

\begin{abstract}
The anisotropic flow of the quark-gluon plasma (QGP) in heavy-ion collisions provides crucial insights into its  transport properties.  Measurements of hadron momentum anisotropies, such as the elliptic flow ($v_2$), have revealed that the QGP  has small viscosity to entropy density ratio and behaves  as a nearly ideal fluid. Unlike cold-atom experiments where temporal evolution of $v_2$ can be directly imaged,  late-stage hadronic  scatterings prevent from resolving the temporal development of the QGP anisotropy. Here, using transport simulations of   $\text{Ru}+\text{Ru}$ and $\text{Zr}+\text{Zr}$  collisions at $\sqrt{s_{NN}} = 200\text{ GeV}$, we show that penetrating dileptons ($e^+ e^-$) can serve as a femtoscale chronoscope for early QGP dynamics. In the intermediate invariant-mass window ($0.8 < M_{ee} < 2.4\text{ GeV}$), we find that dilepton emission is dominated by early QGP radiation within the first $2.0\text{ fm}/c$. Crucially, we find that dilepton $v_2$ scales  with the   partonic $v_2$ while inversely correlating with its temperature. This dual dependence   establishes   mass-differential dilepton measurements as a promising tool for chronometric tomography of the coupled cooling and anisotropic expansion of the QGP  in relativistic heavy-ion collisions. 
\end{abstract}

\date{\today}
\maketitle

\textit{Introduction.}---%
\label{sec:introduction}%
Quantum chromodynamics (QCD), the underlying theory for the strong interaction, predicts a crossover transition at extremely high temperatures ($k_B T\approx 156.5 \pm 1.5$ MeV~\cite{HotQCD:2018pds})  from hadronic matter to quark–gluon plasma (QGP) consisting of deconfined quarks and gluons~\cite{Aoki:2006we}. Such conditions  existed during the first microseconds of the early universe and  have been re-created in ultra-relativistic heavy-ion collisions~\cite{STAR:2005gfr}. The transport properties of QGP can be extracted from measured anisotropic flow coefficients, which quantify the azimuthal expansion of  produced QGP and hadronic matter ~\cite{Heinz:2013th,Teaney:2009qa,Arslandok:2023utm}. In particular, precise measurements of identified hadron flows, combined with  relativistic viscous hydrodynamic modeling of the collisions, have revealed that the QGP behaves as a nearly perfect fluid with a shear viscosity to entropy density ratio near the theoretical lower bound~\cite{Song:2017wtw}. However, unlike cold-atom experiments, where the temporal evolution of elliptic flow ($v_2$) can be directly imaged~\cite{Brandstetter:2023jsy,Li:2024ivj,Yu:2026krf}, hadronic observables in heavy-ion collisions predominantly encode the time-integrated evolution of the expanding system. Consequently, they provide only limited access to the buildup of collective anisotropic flow, particularly during the early stages of QGP evolution.

\begin{figure}[!t]
 \centering
 \includegraphics[width=0.4\textwidth]{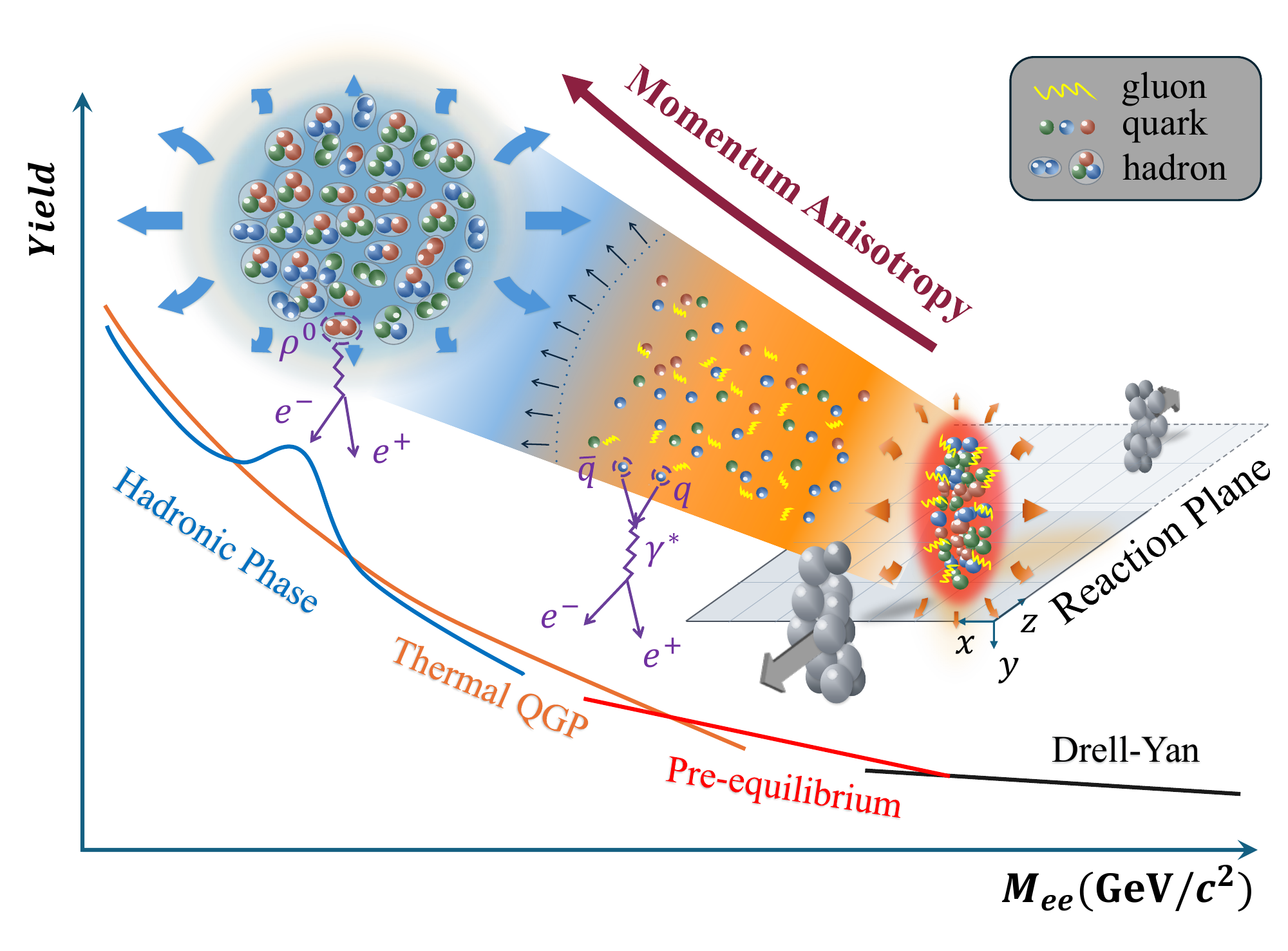}
 \caption{\label{fig:fig0}
Schematic view of dilepton invariant mass spectrum and its production processes in relativistic heavy-ion collisions. The curves sketch the yield originating from in-medium hadronic radiation, mainly the $\rho^0$ (blue), in the low-mass region (LMR); thermal radiation from the deconfined phase (orange) in the intermediate-mass region (IMR); and pre-equilibrium (red) and Drell--Yan (black) at higher masses.  Also shown is the hot fireball  produced in the reaction plane (right), accompanied  by its anisotropic expansion and cooling  (left).}
\end{figure}

The limitation in hadronic observables can be overcome by using  electromagnetic probes.    Both real and virtual photons interact only electromagnetically and therefore escape the expanding medium essentially unscathed  after their production~\cite{Gale:2025emr,Gale:2001in,Rapp:2013nxa}. Dileptons such as $e^+e^-$ pairs from virtual photons are particularly powerful: their invariant mass $M_{ee}$ provides an independent diagnostic of  the emitting source and its characteristic emission epoch~\cite{Rapp:2014hha,HADES:2019auv}. Figure~\ref{fig:fig0} sketches the dilepton  mass-differential spectrum and its production process in relativistic heavy-ion collisions.  In the low-mass region  of $M_{ee}\le1$~GeV, the yield is dominated by in-medium vector mesons, chiefly the $\rho^0$, so the spectrum probes the modification of  their in-medium spectral functions and  other hadronic radiations that continue down to the kinetic freeze-out~\cite{Rapp:1999ej,CERES:1997qfw,NA60:2006rho,NA60:2008thermal,vanHees:2006ng,STAR:2015dielectron,STAR:2014dcy}.  For dileptons in the intermediate-mass region  of $1~{\rm GeV}<M_{ee}<3~{\rm GeV}$,  their invariant-mass spectrum shows a smooth  continuum  due to thermal radiations from  $q\bar q$ annihilation  in the deconfined QGP,  in addition to those from correlated open-charm decays~\cite{Rapp:2014hha,Song:2018plu,Linnyk:2011vx,PHENIX:2015jfz}. At still higher invariant masses of $M_{ee}> 3~{\rm GeV}$, the dilepton spectrum is taken over by pre-equilibrium radiations and  Drell--Yan pairs~\cite{Coquet:2021lca,Drell:1970wh}. Since these processes for dilepton production have different characteristic time scales and contribute to different invariant mass regions, dileptons could serve as a chronometer of the produced matter in relativistic heavy-ion collisions~\cite{Rapp:2014hha,Churchill:2023zkk,Massen:2024pnj}.  Together with direct photons,  these electromagnetic observables could also serve as a tomography of the radial expansion of  produced QGP~\cite{Du:2025dot}. 

In the present study, we exploit dilepton production at different invariant-mass regions as a femtoscale chronoscope to track the temporal development of QGP anisotropic flow.  We employ a multiphase transport (AMPT) framework to investigate dilepton production in high-statistics isobaric ($^{96}_{44}\mathrm{Ru}+{}^{96}_{44}\mathrm{Ru}$ and $^{96}_{40}\mathrm{Zr}+{}^{96}_{40}\mathrm{Zr}$) collisions at $\sqrt{s_{NN}} = 200\text{ GeV}$~\cite{Zhang:2021kxj,STAR:2021mii,STAR:2026isobarv2}. We find that intermediate-mass dileptons ($0.8\text{ GeV} < M_{ee} < 2.4\text{ GeV}$) originate predominantly from early QGP radiation within the first $2.0$~fm/$c$.   Their $v_2$ scales linearly with the underlying quark $v_2$ while exhibiting a distinct inverse correlation with their effective temperature. This dual dependence provides  the signature of simultaneous thermal cooling and momentum anisotropy development, providing a promising tool for the chronometric tomography of early QGP anisotropic dynamics.
\par{}

\begin{figure*}[t]
     \centering
     \includegraphics[width=0.9\textwidth]{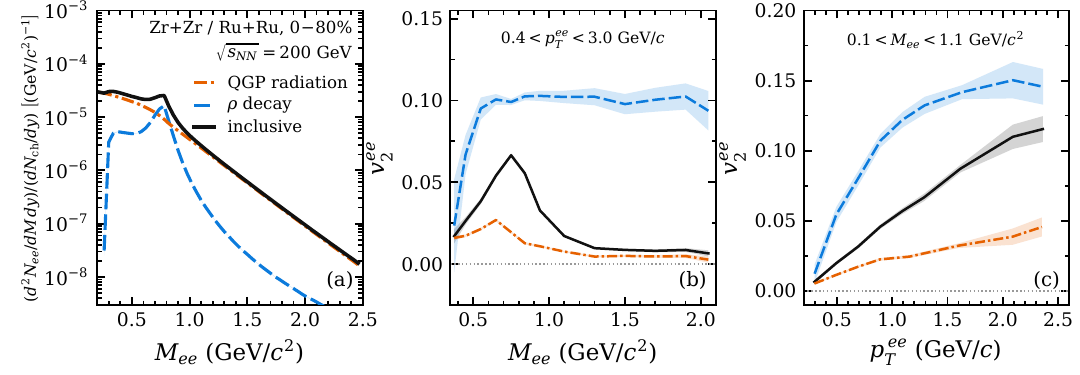}    
\caption{\label{fig:mass_spectrum}  Dilepton invariant-mass spectrum (a),  dilepton elliptic flow $v_2^{ee}$ as a function of $M_{ee}$ for $0.4<p_T^{ee}<3.0$~GeV/$c$ (b), and  elliptic flow as a function of pair transverse momentum for $0.1<M_{ee}<1.1$~GeV/$c^2$ (c).  Shown by orange dash-dotted line is the QGP radiation ($q\bar q\to e^+e^-$), by blue dashed line is the in-medium $\rho$ decay ($\rho^0\to e^+e^-$), and by black solid line is their sum.}
\end{figure*}

\textit{Methods.}---
\label{sec:transport}
We employ  the AMPT model~\cite{Lin:2004en} to describe dilepton production in $^{96}_{44}\text{Ru}+{}^{96}_{44}\text{Ru}$ and $^{96}_{40}\text{Zr}+{}^{96}_{40}\text{Zr}$ collisions at $\sqrt{s_{\mathrm{NN}}}=200\text{~GeV}$ with initial conditions generated  from the HIJING model~\cite{Wang:1991hta,Gyulassy:1994ew} with strings melted to quarks and antiquarks of current masses.  For the  dynamics of these quarks and antiquarks, they evolve under the Nambu--Jona-Lasinio (NJL) mean fields, which  include the scalar and Kobayashi--Maskawa--'t~Hooft (KMT) terms~\cite{Nambu:1961tp,Nambu:1961fr,Buballa:2003qv,Hatsuda:1994pi}. After stopping scatterings, quarks and antiquarks subsequently hadronize via the quark coalescence, and the resulting hadrons undergo rescatterings within a relativistic transport (ART) framework~\cite{Li:1995pra}. 

For  dilepton production, we consider two dominant channels that are known to be important in heavy-ion collisions for the measured ``excess'' of dilepton yields above  those from hadronic decays~\cite{Linnyk:2011vx}.  For the partonic  stage, the dominant channel for dilepton production is  from  quark--antiquark annihilation ($q\bar q\to e^+e^-$) with the leading-order cross section~\cite{Song:2010fk},
\begin{equation}
 \sigma_{q\bar q\to e^+e^-}=\frac{4\pi\alpha^2 e_q^2}{9M^2}\frac{1+2m^2/M^2}{\sqrt{1-4m^2/M^2}},
 \label{eq:qqbar_xs}
\end{equation}
where $m$ is the effective quark mass, $M$ is the invariant mass of the dilepton or quark pair, $e_q$ is the quark charge, and $\alpha=1/137$ is the fine-structure constant. For dilepton production from the hadronic matter, which is nearly baryon free in relativistic heavy-ion collisions, it is mainly from the decay $\rho^0\to e^+e^-$ of neutral $\rho^0$ mesons that are formed  from the quark coalescence  during hadronization and from the reaction $\pi^++\pi^-\to\rho^0$ during the expansion of the hadronic matter~\cite{Li:1994cj,Li:1996mi,Cassing:1999es,Zhou:2025def}. 

 The isospin-averaged cross section for $\rho$ meson production from pion-pion scattering at center-of-mass energy $M$ is~\cite{Li:1996mi}
\begin{equation}
 \sigma_{\pi\pi\to\rho}(M)=
 \frac{8\pi}{k^2}
 \frac{m_\rho^2\,\Gamma_{\rho\to\pi\pi}(M)\,\Gamma_{\rm tot}(M)}
 {(M^2-m_\rho^{2})^2+ \left[m_\rho\Gamma_{\rm tot}(M)\right]^2}
 \left(\frac{M}{m_\rho}\right)^2.
 \label{eq:pipi_rho_xs}
\end{equation}
In the above, $k=\sqrt{M^2/4-m_\pi^2}$ is the pion momentum in the pair center-of-mass frame, $m_\rho=0.775$~GeV is the  vacuum mass of the $\rho$ meson, and $\Gamma_{\rm tot}(M)=\Gamma_{\rho\to\pi\pi}(M)+\Gamma_{\rm coll}$ is its total in-medium width, which includes its decay width $\Gamma_{\rho\to\pi\pi}(M)$ and the medium  induced collisional broadening $\Gamma_{\rm coll}$~\cite{Rapp:1999ej,Bratkovskaya:2007jk}.  For a $\rho$ meson of invariant mass $M$, its decay width to pions is  given by$
 \Gamma_{\rho\to\pi\pi}(M)=
 \frac{g_{\rho\pi\pi}^2}{4\pi}
 \frac{(M^2-4m_\pi^2)^{3/2}}{12M^2}$ 
where $g_{\rho\pi\pi}^2/4\pi=2.84$ and $m_\pi=0.13957\text{~GeV}$. Note that  $\Gamma_{\rho\to\pi\pi}=0$ below the two-pion threshold.

  Following  Ref.~\cite{Bratkovskaya:2007jk}, we parametrize the $\rho$-meson collisional width  as $
 \Gamma_{\rm coll}=\sigma_{\rm eff}\,v_{\rm rel}\,\rho_h$,
where $\rho_h$ is the local density of hadrons, $v_{\rm rel}=\langle|\bm v_\rho-\bm v_j|^2\rangle^{1/2}$ is the root-mean-square relative velocity between the $\rho$ meson and these hadrons, and $\sigma_{\rm eff}$ is  the effective $\rho$-meson absorption cross section. Fitting    the $\rho^0$ absorption rate in ART gives $\sigma_{\rm eff}\approx11.3$~mb for $\rho^0$ mesons formed in $\pi\pi$ collisions and $13.8$~mb for those formed from quark coalescence (See Supplemental Material~\cite{SupplementalMaterial} for more details).

\par{}
For $\rho^0$ meson decay to dileptons, the vector meson dominance gives the mass dependence of the decay width as $ 
 \Gamma_{\rho\to ee}(M)= \Gamma_{\rho\to ee}^{0}\,\left(\frac{m_\rho}{M}\right)^{3}$,
where the coefficient $\Gamma_{\rho\to ee}^{0}=7.04$~keV is determined from the measured $\rho^0\to e^+e^-$ width at $M=m_\rho$~\cite{ParticleDataGroup:2024cfk}. Dilepton production from $\rho^0$ decays can be sampled perturbatively at each ART time step with the probability $1-\exp[-\Delta t\,\Gamma_{\rho\to ee}/\gamma_\rho]$, where $\gamma_\rho$ is the Lorentz factor due to the finite  $\rho^0$ velocity in the hadronic matter~\cite{Li:1994cj,Li:1996mi,Linnyk:2015rco}. Because of the  low dilepton yield, dilepton production from both the partonic and hadronic phases can be treated perturbatively without  including its effect on the dynamics of heavy-ion collisions~\cite{Li:1994cj,Linnyk:2015rco}.

The collision centrality is determined by the charged particle multiplicity at mid pseudo-rapidity, $N_{\rm ch}^{|\eta|<0.5}$. For the elliptic flow $v_2$ of dileptons, it is reconstructed using the two $\eta$-subevent event plane method~\cite{Voloshin:1998mz,STAR:2014dileptonv2}. The second-order flow vectors, $Q_{2,\pm}=\sum_i w_i e^{2i\phi_i}$, in this method are defined using charged hadrons within $0.2<p_T<2\text{~GeV}/c$ and $|\eta|<1$, with a transverse momentum weight $w_i=\min(p_T,2\text{~GeV}/c)$. The events are divided into $\eta_+$ and $\eta_-$ subevents separated by a $\eta$ gap of $|\eta|<0.05$. Each particle of interest is then correlated with the event plane of the opposite subevent, $\Psi_{2,\pm}=\tfrac12\arg Q_{2,\pm}$, i.e., a particle in the $\eta_\pm$ subevent is correlated with $\Psi_{2,\mp}$. For dileptons, $\eta$ is the pseudo-rapidity of the pair, and particles within the gap are not used. The raw flow coefficients are corrected by the event-plane resolution $R_2=\sqrt{\langle\cos2(\Psi_{2,+}-\Psi_{2,-})\rangle}$, resulting in
\begin{equation}
 v_2(\eta_\pm)=\frac{1}{R_2}\,\frac{\sum_{j\in\eta_\pm} W_j\cos2(\phi_j-\Psi_{2,\mp})}{\sum_{j\in\eta_\pm} W_j},
 \label{eq:v2_ep}
\end{equation}
where $W_j=1$ for charged hadrons used in the event-plane definition but is the perturbative production weight for dileptons. The same event plane is used for both hadrons and dileptons. 

\textit{Results.}---We now consider dilepton production in $^{96}_{44}\text{Ru}+{}^{96}_{44}\text{Ru}$ and $^{96}_{40}\text{Zr}+{}^{96}_{40}\text{Zr}$ collisions   in the $0\text{--}80\%$ centrality class at $\sqrt{s_{NN}} = 200\text{ GeV}$.  Since both isobar systems yield almost identical results, we combine their contributions to minimize the statistical uncertainties. Figure~\ref{fig:mass_spectrum}(a) displays the total dilepton invariant-mass spectrum from both  QGP radiation ($q\bar{q} \to e^+e^-$, orange dash-dotted line) and  in-medium rho meson decay ($\rho^0 \to e^+e^-$, blue dashed line). While the QGP radiation  shows a broad continuum across the low- and intermediate-mass regions, the in-medium $\rho^0$ contribution exhibits a pronounced peak  near the $\rho^0$ mass with a value above the contribution from the QGP radiation, resulting in a  total dilepton spectrum  showing a peak near the $\rho^0$ mass  as shown by the black solid line.

The  dilepton elliptic flow $v_2^{ee}$ is  displayed in Fig.~\ref{fig:mass_spectrum}~(b), which shows  that the in-medium $\rho^0$ decay carries a substantially larger $v_2$ than that from the QGP radiation. This is because dileptons  from the QGP radiation escape the system during the early expansion stage of the collisions before the collective flow is fully developed, whereas those from $\rho^0$ decay are emitted after hadronization and up to kinetic freeze-out when the medium has  built up most of its final anisotropy.  

\begin{figure*}[t]
 \centering \includegraphics[width=0.7\textwidth]{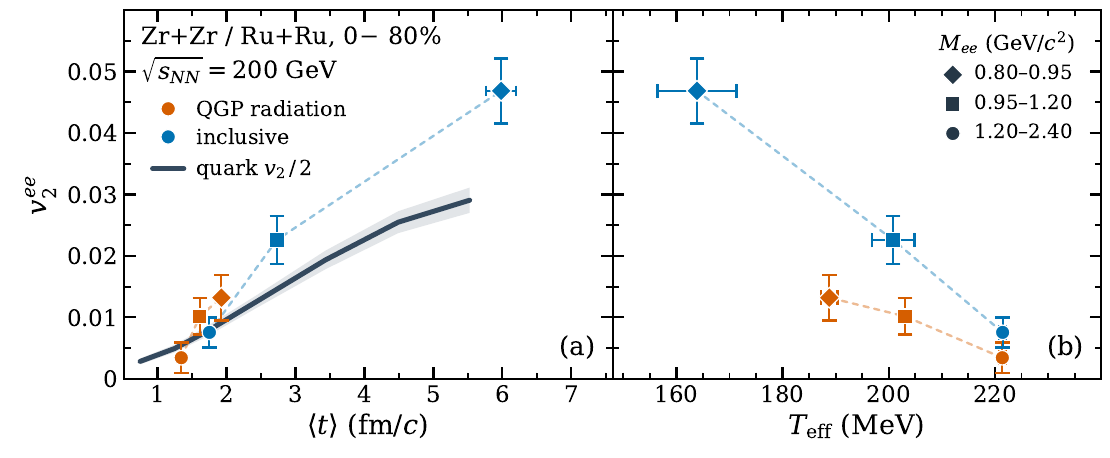}
 \caption{\label{fig:buildup} (a) Dilepton elliptic flow $v_2^{ee}$ versus mean emission time, compared with quark $v_2$ divided by $2$. (b) Dilepton $v_2$ versus the fitted inverse slope $T_{\rm eff}$. Bars and the quark band show jackknife statistical errors; horizontal bars in (a) describe uncertainty of the mean time. QGP radiation and the inclusive signal are shown by orange and blue symbols, respectively. Marker shapes identify the mass windows listed in (b).}
\end{figure*}

Figure~\ref{fig:mass_spectrum} (c) shows the  transverse momentum dependence of  the $v_2^{ee}$ of  invariant mass $0.1<M_{ee}<1.1$~GeV/$c^2$.  It is seen that   $v_2^{ee}$  increases with $p_{T}^{ee}$ for dileptons from both QGP radiation and $\rho^0$ decay, as expected for a collective boost that converts the anisotropy of the flow velocity into a momentum-space anisotropy that grows with $p_T^{ee}$. Similar to the mass dependence of  $v_2^{ee}$,  dileptons from $\rho^0$ decay exhibit a larger $v_2$ than those from QGP radiation, while the yield-weighted average $v_2^{ee}$ lies in between. These results are   consistent with preliminary measurements from the STAR Collaboration~\cite{Bao:2026HP}.

\begin{table}[t] \caption{\label{tab:mass_window} Fraction of the dilepton yield due to QGP radiation $f^{QGP}$, average emission time $\langle t\rangle_{ee}$ and elliptic flow $v_2^{ee}$ of dileptons in three invariant-mass windows of isobaric collisions at $0$--$80\%$ centrality and $0.4<p_T^{ee}<3.0$~GeV/$c$. The upper and lower blocks give the inclusive signal and the QGP radiation, respectively.}
  \begin{ruledtabular}
 \setlength{\tabcolsep}{3pt}
 \begin{tabular}{@{}cccc@{}}
 $M_{ee}$ (GeV/$c^2$) & $f^{\rm QGP}$ & $\langle t\rangle_{ee}$ (fm/$c$)
 & $v_2^{ee}$ \\
 \colrule
 \multicolumn{4}{c}{Inclusive} \\
 $0.80$--$0.95$ & $0.594 \pm 0.005$ & $5.98\pm0.22$
 & $0.047 \pm 0.005$ \\
 $0.95$--$1.20$ & $0.882 \pm 0.002$ & $2.73\pm0.09$
 & $0.023 \pm 0.004$ \\
 $1.20$--$2.40$ & $0.955 \pm 0.001$ & $1.75\pm0.04$
 & $0.008 \pm 0.002$ \\
 \colrule
 \multicolumn{4}{c}{QGP radiation} \\
 $0.80$--$0.95$ &  & $1.93\pm0.06$
 & $0.013 \pm 0.004$ \\
 $0.95$--$1.20$ &  & $1.61\pm0.04$
 & $0.010 \pm 0.003$ \\
 $1.20$--$2.40$ &  & $1.35\pm0.02$
 & $0.003 \pm 0.002$ \\
 \end{tabular}
 \end{ruledtabular}
 \end{table}

For invariant mass $M_{ee} > 0.8\text{ GeV}/c^2$, dilepton production is dominated by QGP radiation, which makes this regime sensitive to partonic dynamics. To trace how the QGP anisotropy develops in time, we analyze dileptons in the three intermediate-mass windows of $0.80\text{--}0.95$, $0.95\text{--}1.20$, and $1.20\text{--}2.40\text{ GeV}/c^2$. Table~\ref{tab:mass_window} lists, for each window, the fraction $f^{\rm QGP}$ of dilepton yield due to QGP radiation, together with the average emission time and the elliptic flow of the dileptons for both the inclusive signal and  the QGP radiation alone.  The mean dilepton emission time is denoted as $\langle t\rangle_{ee}$.  With increasing dilepton invariant mass, $f^{\rm QGP}$  is seen to increase from $0.59$ to $0.95$, while for the QGP radiation the emission time $\langle t\rangle_{ee}$ decreases from $1.98$ to $1.37$~fm/$c$ and  its flow $v_2^{ee}$ decreases from $0.0117$ to $0.0037$. Dilepton of higher invariant mass therefore selects radiation at earlier times and carries a smaller anisotropy. This  behavior follows from the buildup of the QGP momentum anisotropy during  its expansion and  plays a key  role in chronoscoping the anisotropic expansion of the QGP.

Figure~\ref{fig:buildup}(a) shows the time evolution of the elliptic flow  for dileptons and quarks (solid curve), where the quark $v_2$ is divided by 2 for comparison. The scaled quark $v_2$ closely follows the dilepton elliptic flow from quark-gluon plasma (QGP) radiation, indicating that mass-selected dileptons accurately trace the real-time buildup of partonic momentum anisotropy and serve as a reliable chronometer of early QGP dynamics. Note that this scaling differs from quark number scaling in conventional quark coalescence, where nearly collinear quarks combine such that the meson $v_2$ is roughly twice the quark $v_2$. By contrast, for dileptons with $M_{ee} > 0.8\text{ GeV}$ produced via quark-antiquark annihilation, the average opening angle is large ($\approx 75^\circ$), resulting in a smaller dilepton $v_2^{ee}$ than that of the constituent quarks.

During the QGP expansion, the buildup of anisotropic flow  is accompanied by a cooling process. For instance, the HADES Collaboration  has extracted a temperature of $71.8\pm2.1$~MeV from measured dilepton  invariant mass spectrum in Au+Au collisions at $\sqrt{s_{NN}}=2.42$~GeV~\cite{HADES:2019auv}, and the STAR Collaboration has obtained a higher temperature of $287\pm70(\text{stat.})\pm34(\text{syst.})$~MeV for the hot near baryon-free matter in collisions at $\sqrt{s_{NN}}=54.4$~GeV~\cite{STAR:2024bpc}. 

The effective dilepton temperature $T_{\text{eff}}$ extracted from HADES and STAR experiments are based on fitting the dilepton invariant-mass spectrum by the  parameterization~\cite{Geurts:2022xmk}
\begin{equation}
 \label{eq:teff_spectrum}
 \frac{dN_{ee}}{dM} = A \left(M T_{\rm eff}\right)^{3/2} \exp\left(-\frac{M}{T_{\rm eff}}\right),
 \end{equation}
 where $A$ is a normalization constant.  Equation~(\ref{eq:teff_spectrum}) describes a smooth continuum and applies wherever the QGP radiation dominates.  Since the inclusive dilepton invariant-mass spectrum below $M_{ee}\approx1.2$~GeV/$c^2$  contains the in-medium $\rho^0$, whose resonant lineshape is not exponential,  a single-component fit  using Eq.~(\ref{eq:teff_spectrum}) only gives information about the  composition of the dilepton emission source rather than  the slope of the dilepton invariant-mass spectrum. For the two low-mass windows of the inclusive  dileptons, we therefore add a hadronic term   that has a similar spectral  form to that from the transport  model simulations~\cite{Jin:2025tke}: 
\begin{equation}
 \label{eq:teff_lmr}
\frac{dN_{ee}}{dM} = \left[\,a\left(M T_{\rm eff}\right)^{3/2}
 + b\,S_\rho(M)\right]
 \exp\left(-\frac{M}{T_{\rm eff}}\right),
\end{equation}
 where $a$, $b$ and $T_{\rm eff}$ are parameters and 
\begin{equation}
 \label{eq:s_rho}
S_\rho(M)=\dfrac{\left[m_\rho\Gamma_{\rho\to\pi\pi}(M)\right]^2} {(M^2-m_\rho^{2})^2+\left[m_\rho\Gamma_{\rho\to\pi\pi}(M)\right]^2}\,\Gamma_{\rho\to ee}(M)
\end{equation}
For the $1.20\text{--}2.40$~GeV/$c^2$ invariant-mass window, where the $\rho^0$  contributes only $4.4\%$ of the dilepton yield, Eq.~(\ref{eq:teff_spectrum}) is thus a good approximation. See the Supplemental Material ~\cite{SupplementalMaterial} for more details on extracting the effective temperature by fitting dilepton spectrum.

Figure~\ref{fig:buildup}~(b) plots dilepton $v_2^{ee}$ against $T_{\text{eff}}$. The highest-mass window ($1.20\text{--}2.40\text{ GeV}/c^2$) yields the highest inverse slope parameter ($T_{\text{eff}} = 221.4 \pm 0.6\text{ MeV}$) and the smallest anisotropy ($v_{2,ee}^{\rm QGP} = 0.003$). For lower-mass windows, $T_{\text{eff}}$ steadily decreases to $203.1 \pm 0.9\text{ MeV}$ and $188.8 \pm 1.6\text{ MeV}$, while $v_2^{ee}$ from QGP radiation increases more than two-fold. This  anti-correlation between $v_2^{ee}$ and $T_{\text{eff}}$ provides a signature of simultaneous thermal cooling and anisotropic expansion of the partonic matter.

In experimental measurements,  dileptons from the $\rho^0$ decay and the QGP radiation are difficult to separate. The blue symbols  in Fig.~\ref{fig:buildup} show results for dileptons from  both QGP radiation and $\rho$ decay.   The inclusive $v_2^{ee}$ exceeds that of the QGP radiation by factors of $2.2$ to $3.5$, and the inclusive emission time increases from $1.75$ to $5.98$~fm/$c$, since the $\rho^0$ continues to decay until kinetic freeze-out. For the inclusive dilepton spectrum in panel (b) of Fig.~\ref{fig:buildup}, the inverse slope of the two low-mass windows is obtained from Eq.~(\ref{eq:teff_lmr}) and that of the highest window from Eq.~(\ref{eq:teff_spectrum}).  The effective temperature is seen to decrease monotonically from $221.5\pm0.6$ to $200.8\pm4.0$ and $163.9\pm7.4$~MeV, with the lowest temperature  close to the kinetic freeze-out temperature. Results shown in Fig.~\ref{fig:buildup} suggest that dileptons serve as a feasible tool for the chronometric tomography of strongly-interacting matter in relativistic heavy-ion collisions.  

The extracted effective temperature in Fig.~\ref{fig:buildup} is consistent with hydrodynamic calculations~\cite{Churchill:2023zkk,Zhang:2025kjr}, but is slightly lower than that extracted from preliminary STAR data~\cite{STAR:2024bpc,Bao:2026HP}. This discrepancy may arise from contributions of dileptons produced during the pre-equilibrium stage of heavy-ion collisions, as well as contamination from multi-pion decays~\cite{vanHees:2007th,Endres:2014zua}, which are difficult to disentangle and subtract from the dilepton spectra in experimental analyses.

\textit{Summary.}---%
 \label{sec:summary}%
 In summary, we have demonstrated that mass-differential dilepton emission serves as a femtoscale chronoscope for mapping the temporal evolution of quark-gluon plasma (QGP) anisotropic expansion in relativistic heavy-ion collisions, analogous to cold-atom experiments. Using a multiphase transport framework to analyze isobaric ($\mathrm{Ru}+\mathrm{Ru}$ and $\mathrm{Zr}+\mathrm{Zr}$) collisions at $\sqrt{s_{NN}} = 200\text{ GeV}$, we show that intermediate-mass dileptons ($0.8\text{ GeV} < M_{ee} < 2.4\text{ GeV}$) predominantly probe partonic radiation with mean emission times of $1.35\text{--}1.9\text{ fm}/c$. Specifically, the dilepton elliptic flow $v_2^{ee}$ in this mass window directly tracks the buildup of bulk partonic anisotropy while exhibiting a clear inverse correlation with the dilepton effective  temperature $T_{\text{eff}}$. Notably, this anticorrelation persists even when hadronic contributions from $\rho$-meson decays are included alongside QGP radiation, providing an unambiguous signature of simultaneous thermal cooling and anisotropic expansion. Mass-differential dilepton tomography thus offers a promising chronometric probe of early-stage QGP dynamics across current RHIC and LHC programs as well as future facilities such as FAIR and NICA.

{\it Acknowledgments.--}
The authors thank helpful discussions with Wang-Mei Zha, Zao-Chen Ye, Shan-Shan Cao. This work was supported in part by the National Natural Science Foundation of China under contract No. 12422509, No. 12375121, No. 12147102, No.12547102, No. 12405150, No. 12347106, No. 12347143, and No. 12688101; and the National Key Research and Development Project of China under Grant No. 2024YFA1612500. The computations in this research were performed using the CFFF platform of Fudan University. 


%

\clearpage
\onecolumngrid
\setcounter{equation}{0}
\setcounter{figure}{0}
\setcounter{table}{0}
\setcounter{section}{0}
\setcounter{secnumdepth}{2}
\renewcommand{\theequation}{S\arabic{equation}}
\renewcommand{\thefigure}{S\arabic{figure}}
\renewcommand{\thetable}{S\Roman{table}}
\renewcommand{\thesection}{\Roman{section}}
\renewcommand{\theHequation}{S\arabic{equation}}
\renewcommand{\theHfigure}{S\arabic{figure}}
\renewcommand{\theHtable}{S\arabic{table}}
\renewcommand{\theHsection}{S\arabic{section}}

\begin{center}
\textbf{\large Supplemental Material for ``Chronoscoping Anisotropic Flow in the Quark-Gluon Plasma with Dileptons''}\\[6pt]
\end{center}
\vspace{6pt} 
\section{Collisional broadening width  of \texorpdfstring{$\rho^0$}{rho0} meson }
\label{sec:supp_width}

We use natural units ($\hbar=c=1$, with $1~\text{mb}=0.1~\text{fm}^2$) throughout the following sections. In the main text, the in-medium total width of the $\rho^0$ meson, $\Gamma_{\rm tot}(M)=\Gamma_{\rho\to\pi\pi}(M)+\Gamma_{\rm coll}$, governs both the $\pi^+\pi^-\to\rho^0$ formation cross section in Eq.~(\ref{eq:pipi_rho_xs}) and the spectral mass sampling during quark coalescence. While retaining the vacuum pole mass $m_\rho=0.775$~GeV/$c^2$ and vacuum strong decay width $\Gamma_{\rho\to\pi\pi}(M)$, we parameterize the collisional broadening width as $\Gamma_{\rm coll}=\sigma_{\rm eff}\,v_{\rm rel}\,\rho_h$~\cite{Rapp:1999ej,Bratkovskaya:2007jk} and extract the effective absorption cross section $\sigma_{\rm eff}$ from microscopic $\rho^0$ absorption rates in the ART hadronic cascade.

Specifically, the local hadron density $\rho_h$ and root-mean-square relative velocity $v_{\rm rel}$ are evaluated on a cubic spatial lattice of cell volume $V_c=1$~fm$^3$ in the collision center-of-mass frame:
\begin{equation}
 \rho_h=\frac{1}{V_c}\sum_j w_j,\qquad
 v_{\rm rel}^2=\frac{\sum_j w_j|\bm v_\rho-\bm v_j|^2}{\sum_j w_j},
 \qquad \bm v_j=\frac{\bm p_j}{E_j}.
 \label{eq:supp_local}
\end{equation}
To suppress lattice fluctuations, each surrounding hadron is smeared across neighboring cells with weight $1/3$ in its resident cell and $1/9$ in each of the six nearest-neighbor cells; equivalently, a target cell sums surrounding hadrons inside itself with $w_j=1/3$ and those in its six adjacent cells with $w_j=1/9$. Both the $\rho^0$ meson under consideration and all pions are excluded from Eq.~(\ref{eq:supp_local}). Although pions account for roughly $41\%$ of the surrounding hadron density in our simulations, they primarily undergo elastic scatterings or $\rho$ regeneration and contribute only $\sim 4.5\%$ to net $\rho^0$ absorption. The effective cross section $\sigma_{\rm eff}$ is therefore defined relative to non-pion hadrons. For $\pi^+\pi^-\to\rho^0$ formation, Eq.~(\ref{eq:supp_local}) is evaluated at the midpoint of the pion pair with $\bm v_\rho=(\bm p_1+\bm p_2)/(E_1+E_2)$. For $\rho^0$ mesons formed via quark coalescence, invariant masses are sampled over $2m_\pi\le M\le 3.0$~GeV/$c^2$ from the relativistic Breit-Wigner distribution
\begin{equation}
 P(M)\propto
 \frac{[m_\rho\Gamma_{\rm tot}(M)]^2}
 {(M^2-m_\rho^2)^2+[m_\rho\Gamma_{\rm tot}(M)]^2},
 \label{eq:supp_mass}
\end{equation}
evaluated at the coalescence position; at fixed three-momentum $\bm p_\rho$, the velocity $\bm v_\rho(M)=\bm p_\rho/\sqrt{\bm p_\rho^2+M^2}$, and hence $v_{\rm rel}$ and $\Gamma_{\rm coll}(M)$, depends on $M$.

To calibrate $\sigma_{\rm eff}$, we perform AMPT simulations of Ru+Ru collisions at $\sqrt{s_{NN}}=200$~GeV (0-80\% centrality, time step $\Delta t=0.2$~fm/$c$) with collisional broadening turned off ($\Gamma_{\rm coll}=0$) in both the $\pi^+\pi^-\to\rho^0$ cross section and the coalescence mass distribution, while keeping all strong decays and inelastic hadronic channels active. We track $\rho^0$ mesons produced via $\pi\pi$ formation and quark coalescence within $0.28\le M<1.2$~GeV/$c^2$ and identify net absorption events as hadronic reactions in which no $\rho$ meson survives in the final state, excluding strong decays, charge exchange, and elastic scatterings. Equating the observed absorption count $N_{\rm abs}$ to the path-integrated absorption rate along the tracked $\rho^0$ trajectories yields
\begin{equation}
 \sigma_{\rm eff}=
 \frac{N_{\rm abs}}
 {\sum_i\Delta\tau_i\,v_{{\rm rel},i}\,\rho_{h,i}},
 \label{eq:supp_fit}
\end{equation}
where the sum runs over all tracked $\rho^0$ trajectory segments of proper-time duration $\Delta\tau_i=\Delta t/\gamma_i$, with Lorentz factor $\gamma_i$. Evaluating Eq.~(\ref{eq:supp_fit}) separately for the two production mechanisms gives $\sigma_{\rm eff}\approx11.3$~mb for $\pi\pi$ formation and $13.8$~mb for quark coalescence.

\section{Invariant-mass spectral fitting}
\label{sec:supp_teff}
The effective temperatures  $T_{\rm eff}$ in Fig.~\ref{fig:buildup}(b) of the main text are extracted from the charged-multiplicity-normalized invariant-mass spectrum $y(M)=(d^2N_{ee}/dM\,dy)/(dN_{\rm ch}/dy)$ within $0.4<p_T^{ee}<3.0$~GeV/$c$ and $|y_{ee}|<1$ across $15\,606$ combined isobar ($\text{Ru}+\text{Ru}$ and $\text{Zr}+\text{Zr}$) events at $0$--$80\%$ centrality. In each mass window, the spectrum is binned into $N=15$ mass intervals of center $M_k$ and fitted with Eqs.~(\ref{eq:teff_spectrum})--(\ref{eq:s_rho}) of the main text, with $M$ and $T_{\rm eff}$ expressed in GeV so that the normalization coefficients $A$ and $a$ have units of GeV$^{-4}$ and $b$ has units of GeV$^{-2}$. Specifically, the single-component continuum [main-text Eq.~(\ref{eq:teff_spectrum})] is fitted to pure QGP radiation in all three windows and to the inclusive spectrum in $1.20$--$2.40$~GeV/$c^2$, where the $\rho^0$ contribution is only $4.5\%$. For the inclusive spectra in $0.80$--$0.95$ and $0.95$--$1.20$~GeV/$c^2$, we apply the two-component form [main-text Eq.~(\ref{eq:teff_lmr})] with the vacuum $\rho^0$ spectral template $S_\rho(M)$ [main-text Eq.~(\ref{eq:s_rho})] and $m_\rho=0.775$~GeV/$c^2$.

Model parameters are obtained by minimizing the weighted $\chi^2$,
\begin{equation}
 \chi^2=\sum_{k=1}^{N}\frac{\left[y_k-F(M_k)\right]^2}{\sigma_k^2},
 \label{eq:supp_chi2}
\end{equation}
where $y_k$ is the normalized differential yield in bin $k$, $\sigma_k$ is its jackknife statistical uncertainty estimated across $100$ independent event subsamples (with $dN_{\rm ch}/dy$ propagated consistently), and $F(M_k)$ denotes Eq.~(\ref{eq:teff_spectrum}) or Eq.~(\ref{eq:teff_lmr}) of the main text. Statistical uncertainties on $T_{\rm eff}$ and the normalization coefficients ($A$, $a$, and $b$) are evaluated by repeating the $\chi^2$ minimization for each leave-one-out jackknife sample with the bin weights $1/\sigma_k^2$ held fixed, thereby preserving bin-to-bin statistical correlations.

\begin{figure}[!t]
 \centering
 \includegraphics[width=0.88\linewidth]{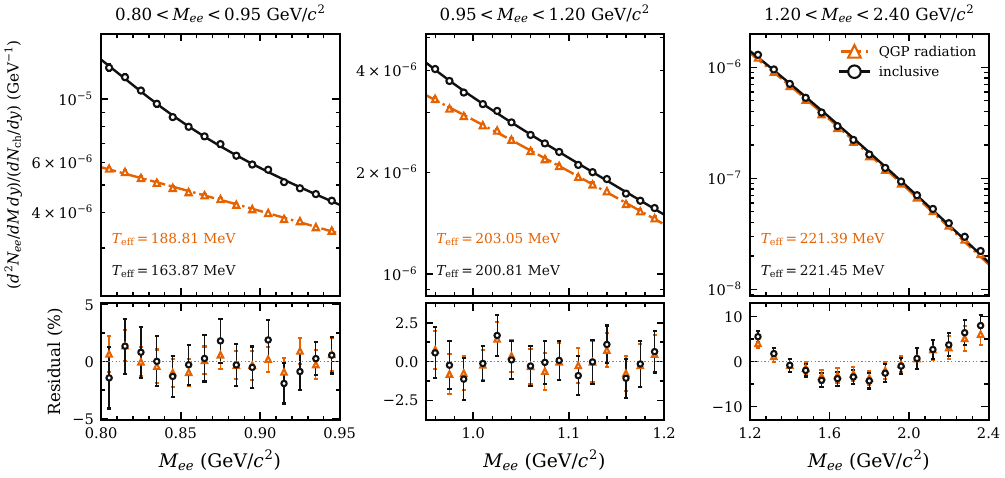}
 \caption{\label{fig:supp_window}
 Upper panels: normalized dilepton invariant-mass spectra and weighted $\chi^2$ fits across the three mass windows analyzed in the main text. Pure QGP radiation (orange triangles) is fitted with Eq.~(\ref{eq:teff_spectrum}) of the main text, while the inclusive spectrum (black circles) is fitted with Eq.~(\ref{eq:teff_lmr}) in the two lower-mass windows and Eq.~(\ref{eq:teff_spectrum}) in the highest-mass window.
 Lower panels: relative fit residuals $100[y_k/F(M_k)-1]$. Error bars show jackknife statistical uncertainties $\sigma_k$ of the simulated spectra. The extracted effective temperatures $T_{\rm eff}$ are labeled in colors matching the corresponding spectra.}
\end{figure}

Figure~\ref{fig:supp_window} compares the simulated invariant-mass spectra with the weighted $\chi^2$ fits and relative residuals, and Table~\ref{tab:supp_window} lists all fitted parameters together with $\chi^2/\mathrm{ndf}$. In the two lower-mass windows, $\chi^2/\mathrm{ndf}<1$ reflects positive bin-to-bin yield correlations omitted from the diagonal $\chi^2$, which are accounted for in the quoted uncertainties via the subsample jackknife. Adopting the fixed vacuum $\rho^0$ template $S_\rho(M)$ follows experimental extractions in which the in-medium collisional broadening is \textit{a priori} unknown; part of the broadened hadronic tail is absorbed into the continuum term, so the low-mass inclusive $T_{\rm eff}$ values are defined relative to this vacuum template. In the widest window ($1.20$--$2.40$~GeV/$c^2$), sub-percent statistical errors resolve a mild positive curvature in the relative residuals ($\chi^2/\mathrm{ndf}>1$) arising from the superposition of thermal emission over continuously cooling QGP stages, together with the $4.5\%$ residual $\rho^0$ tail in the inclusive spectrum.

\begin{table}[!b]
 \caption{\label{tab:supp_window}
 Summary of weighted $\chi^2$ spectral fit parameters across the three invariant-mass windows for $0.4<p_T^{ee}<3.0$~GeV/$c$ and $|y_{ee}|<1$. The fourth column reports $A$ for single-component continuum fits [main-text Eq.~(\ref{eq:teff_spectrum})] and $a$ for two-component fits [main-text Eq.~(\ref{eq:teff_lmr})]. Uncertainties are jackknife statistical errors; a dash denotes an absent hadronic term.}
 \footnotesize
 \setlength{\tabcolsep}{5pt}
 \begin{ruledtabular}
 \begin{tabular}{@{}llcccc@{}}
 $M_{ee}$ (GeV/$c^2$) & Sample & $T_{\rm eff}$ (MeV)
 & $A$ or $a$ ($10^{-3}$~GeV$^{-4}$) & $b$ (GeV$^{-2}$) & $\chi^2/\mathrm{ndf}$ \\
 \colrule
 $0.80$--$0.95$ & QGP radiation & $188.81\pm1.59$ & $6.80\pm0.37$ & --- & $4.52/13$ \\
  & inclusive & $163.87\pm7.40$ & $21.05\pm7.00$ & $140.4\pm22.0$ & $5.08/12$ \\
 \colrule
 $0.95$--$1.20$ & QGP radiation & $203.05\pm0.88$ & $4.30\pm0.13$ & --- & $4.63/13$ \\
  & inclusive & $200.81\pm4.01$ & $4.91\pm0.71$ & $90.9\pm9.1$ & $4.76/12$ \\
 \colrule
 $1.20$--$2.40$ & QGP radiation & $221.39\pm0.62$ & $2.22\pm0.03$ & --- & $46.25/13$ \\
  & inclusive & $221.45\pm0.59$ & $2.31\pm0.03$ & --- & $72.24/13$ \\
 \end{tabular}
 \end{ruledtabular}
\end{table}

\end{document}